\documentclass[useAMS,usenatbib,usegraphicx]{mn2e}
\usepackage{graphicx,times,epsfig,rotating}

\def\mnras{MNRAS}
\def\aj{AJ}
\def\aap{A\&A}
\def\apj{ApJ}
\def\apjl{ApJ}
\def\apjs{ApJS}
\def\araa{ARA\&A}
\def\pasp{PASP}
\def\nat{Nature}

\def\aaps{A\&AS}

 \newcommand{\kms}{\mbox{km s$^{-1}$ }}

\newcommand{\hpc}{\mbox{h$^{-1}$pc }}

\newcommand{\msun}{\mbox{M$_{\sun}$ }}
\newcommand{\msunend}{\mbox{M$_{\sun}$}}

\newcommand{\cmthree}{\mbox{cm$^{-3}$}}

\newcommand{\msunyr}{\mbox{M$_{\sun}$yr$^{-1}$ }}
\newcommand{\msunyrend}{\mbox{M$_{\sun}$yr$^{-1}$}}
\newcommand{\htwo}{\mbox{H$_2$}}

\newcommand{\z}{\mbox{$z$}}

\newcommand{\zsim}{\mbox{$z\sim$ }}

\newcommand{\sunrise}{\mbox{\sc sunrise}}
\newcommand{\gadget}{\mbox{\sc gadget-3}}
\newcommand{\starburst}{\mbox{\sc starburst99}}
\newcommand{\mappings}{\mbox{\sc mappingsiii}}

\newcommand{\mc}{\mbox{$\hat{M_{\rm c}}$}}
\newcommand{\msunpcsq}{\mbox{$\msun {\rm pc}^{-2}$}}

\newcommand{\fsps}{\mbox{\sc fsps}}

\newcommand{\gadgettwo}{\mbox{\sc gadget-2}}

\title[Cosmic Evolution of IMF and Bottom-Heavy Ellipticals]{The Cosmic Evolution of the IMF Under the Jeans Conjecture with Implications for Massive Galaxies}

\author[Narayanan \& Dav\'e]{Desika\, Narayanan$^{1,5}$\thanks{E-mail:
    dnarayanan@as.arizona.edu}\thanks{Bart J Bok Fellow} \& Romeel
  Dav\'e$^{2,3,4,5}$\\$^{1}$Department of Astronomy, Haverford
  College, 370 Lancaster Ave, Haverford, PA, 19041\\$^{2}$University
  of the Western Cape, 7535 Bellville, Cape Town, South
  Africa\\$^{3}$South African Astronomical Observatories, Observatory,
  Cape Town, 7925, South Africa\\$^{4}$African Institute for
  Mathematical Sciences, Muizenberg, Cape Town, 7945, South Africa
  \\$^{5}$Steward Observatory, University of Arizona, 933 N Cherry
  Ave, Tucson, Az, 85721}
\begin{document}

\date{Submitted to MNRAS}

\pagerange{\pageref{firstpage}--\pageref{lastpage}} \pubyear{2010}

\maketitle

\label{firstpage}

\begin{abstract}

We examine the cosmic evolution of a stellar initial mass function
(IMF) in galaxies that varies with the Jeans mass in the interstellar
medium, paying particular attention to the $K$-band stellar mass to
light ratio ($M/L_K$) of present-epoch massive galaxies.  We calculate
the typical Jeans mass using high-resolution hydrodynamic simulations
coupled with a fully radiative model for the ISM, which yields a
parameterisation of the IMF characteristic mass as a function of
galaxy star formation rate (SFR).  We then calculate the star
formation histories of galaxies utilising an equilibrium galaxy growth
model coupled with constraints on the star formation histories set by
abundance matching models.  We find that at early times,
energetic coupling between dust and gas drive warm conditions in the
ISM, yielding bottom-light/top-heavy IMFs associated with large ISM
Jeans masses for massive star-forming galaxies.  Owing to the remnants
of massive stars that formed during the top-heavy phases at early
times, the resultant $M/L_K(\sigma)$ in massive galaxies at the
present epoch is {\it increased} relative to the non-varying IMF case.
At late times, lower cosmic ray fluxes allow for cooler ISM
temperatures in massive galaxies, and hence newly formed clusters will
exhibit bottom-heavy IMFs, further increasing $M/L_K(\sigma)$.  Our 
central result is hence that a given massive galaxy may go
through both top-heavy and bottom-heavy IMF phases during its lifetime,
though the bulk of the stars form during a top-heavy phase.
Qualitatively, the variations in $M/L_K(\sigma)$ with
galaxy mass are in agreement with observations, however, our
model may not be able to account for bottom-heavy mass functions
as indicated by stellar absorption features.

\end{abstract}

\section{Introduction}
\label{section:introduction}
The stellar initial mass function (IMF) is fundamental for
understanding a wide range of problems in astrophysics.  The IMF
provides a mechanism for parameterising the distribution of masses of
stars formed in a given generation.  Beyond being a crucial ingredient
for any successful theory of star formation \citep[e.g.][]{mck07}, the
IMF determines such galactic properties as the metal enrichment
history, energy distribution of photons injected into the interstellar
medium (ISM), and conversion of luminosities into physical quantities
such as star formation rates (SFRs) and stellar masses ($M_*$).

Despite the obvious importance of the IMF, no consensus exists that
describes its origin and possible variations. In the Milky Way, direct
star counts have found that the IMF may be well parameterised by a
single power-law \citep[][]{sal55}, a multi-segment powerlaw
\citep[e.g.][ hereafter, ``Kroupa-like'' IMFs]{kro93,kro02}, or a
lognormal \citep[e.g.][]{mil79,cha03d}.  Generally, observations have
found at most mild variations in the IMF within the Galaxy \citep{bas10}.

From a theoretical standpoint, a broad range of models for the origin
of the IMF exist.  The central ideas behind these range from giant
molecular cloud (GMC) Jeans mass arguments
\citep{lar98,lar05,bat05,tum07,elm08,nar12b} to protostellar feedback
models \citep{sil95,ada96,bat09,kru11d} to models of a turbulent ISM
\citep{pad02,hen08,hop12b}.  While the driving physics behind the
origin of the IMF varies substantially, nearly all theories suggest
that the IMF should vary with changing physical conditions, even if
relatively weakly \citep[c.f.][]{elm08,kru11d}. Though there is
currently no unambiguous direct evidence for systematic variations of
the IMF with environmental conditions \citep{bas10}, the last decade
of observations have provided tantalizing {\it indirect} evidence for
two emerging trends in IMF variations.

The first suggests that the IMF may be bottom-heavy (defined as an
excess of the ratio of low mass stars to high mass stars with respect
to a Milky Way IMF) in present-epoch early-type galaxies. Complementary
observational techniques have arrived at this conclusion, adding
to its robustness.  For example, recent observations of gravity-sensitive
stellar absorption lines (such as FeH -- the ``Wing-Ford'' band --
Ca II, and Na I among others) combined with new empirical stellar
libraries of cool stars \citep{ray09} have been able to distinguish
K and M dwarfs from K and M giants, and found the IMF to be
bottom-heavy in \zsim 0 early-type galaxies
\citep[][]{win69,van10,van12,con12a,con12b,spi12,fer12,smi12}.
Meanwhile, observed constraints on the stellar mass to light ratio
from stellar kinematics have arrived at the same conclusion
\citep[e.g.][]{aug10,tre10,spi11,cap12b,cap12,bre12,dut12,tor12}.
Interestingly, both methods find that the IMF varies systematically
in a manner that more massive galaxies have increasingly bottom-heavy
IMFs.

If elliptical galaxies are the end-product of massive starbursts, then
the implication is that the IMF may show an excess of low-mass stars
in heavily star-forming environments. This is, however, in direct
contrast to a number of indirect measurements of the IMF in starbursts
which all argue that the IMF may be {\it top-heavy} in these
environments, rather than bottom-heavy.  Evidence for this second
trend in IMF variations comes from a variety of complementary
albeit indirect measurements.

\citet{tac08} performed simultaneous modeling of the
stellar masses, CO-\htwo \ conversion factors and stellar initial mass
functions of \zsim 2 Submillimetre-selected galaxies (SMGs), the most
rapidly star-forming galaxies in the Universe, and found a best fit
model with a mass to light ratio roughly half that of a standard
Kroupa IMF.  Similarly, \citet{van08} suggested that the colour
evolution and mass to light ratios for \zsim 1 early-type galaxies may
best be explained by a bottom-light IMF.  More indirect evidence comes
from models that have appealed to a top-heavy IMF at \zsim 2 to
resolve the noted discrepancy between the observed SFR-$M_*$ relation
and simulated one \citep{dav08}.

Some evidence for a top-heavy IMF in heavily star-forming systems at
\zsim 0 exists as well.  For example, \citet{rie93} and \citet{for03}
suggest that the IMF may have a turnover mass a factor $\sim2-6$
larger than traditional \citet{kro02} IMF in the nearby starburst
galaxy M82 \citep[though see][]{sat97}.  Similarly, a simultaneous fit
to the present day $K$-band luminosity density, observed cosmic star
formation rate density and cosmic background radiation by
\citet{far07} suggest a ``paunchy'' IMF wherein there is an excess of
intermediate mass stars.  Utilising a sample of $\sim 33,000$
galaxies, \citet{gun11} find evidence for a strong IMF-SFR relation
in galaxies, such that heavily star forming systems appear to have a
more top-heavy IMF.  Even in the Galactic Centre, observations suggest
that a top-heavy IMF may apply \citep{nay05,sto05}.  While all of
these observations can be explained without the need for IMF
variations\footnote{In \S~\ref{section:discussion}, we review many of
  the arguments against a varying IMF in the context of the
  aforementioned examples.}, it is compelling that when a variable IMF
is inferred for a high-star formation rate surface density
environment, it invariably tends toward top-heavy.

In short, while the evidence is by no means firm, deviations from a
Milky Way IMF appear to follow two trends: in present-day ellipticals,
the IMF tends toward an excess of low-mass stars, whereas in heavily
star-forming galaxies from \zsim 0-2, the IMF appears to have an
excess of high-mass stars.  Simultaneously understanding the origin of
both of these trends presents a challenge for any theory of the IMF.

One possibility is that systematic variations in the IMF owe to
varying physical conditions in the star-forming molecular ISM.
\citet{jea02} and others \citep[e.g.][]{lar05,bat05,kle07,tum07,nar12b}
have postulated that the characteristic stellar mass formed may
relate to the fragmentation properties of the parent molecular cloud
(this is commonly known as the ``Jeans Conjecture'').  In this
picture, the temperature and density of the molecular cloud ($M_J
\sim T^{3/2}/n^{1/2}$) determine the minimum fragmentation mass of
the cloud, and thus set the stellar mass scale.

In this paper, we explore the cosmic evolution of the IMF in galaxies
under the Jeans Conjecture to examine whether a stellar IMF that
varies with the Jeans masses of molecular clouds can explain the
tentative evidence for bottom-heavy IMFs in early-type galaxies as
well as the purported top-heavy IMF in high-star formation rate
density galaxies.  It is important to note that we neither argue for
or against the aforementioned observed IMF trends, but rather take
them as a given.  Similarly, we make no claims regarding the validity
of the Jeans Conjecture, but rather seek to assess its plausibility by
examining the cosmological consequences of a stellar IMF that varies
with the Jeans mass in molecular clouds.  In \S~\ref{section:methods},
we describe our methods, detailing our galaxy formation models as well
as model for the molecular ISM.  In \S~\ref{section:results}, we
report our main results, and in \S~\ref{section:discussion}, we
provide discussion.  Finally, in \S~\ref{section:summary}, we
summarise.

\section{Galaxy Evolution and ISM Models}
\label{section:methods}
\subsection{Summary of Methods}
Our principle goal in this work is to model the cosmic evolution of
the IMF under the assumption that the IMF varies with the Jeans
properties of molecular clouds.  Because resolving molecular cloud
scales ($\sim 10$s of pc) in cosmological volumes is intractable for
all but a few galaxies \citep[e.g.][]{chr12}, we have developed a
multi-scale methodology in order to model the mean ISM properties for
an ensemble of galaxies.

Our first goal is to parameterise the mean molecular ISM physical
properties in terms of a global property.  To do this, we employ a
large suite of relatively high-resolution smoothed particle
hydrodynamic (SPH) simulations of isolated disc galaxies and galaxy
mergers in evolution.  These galaxies sample a range of galaxy masses
and merger mass ratios, with the main goal being to simulate a large
dynamic range of physical conditions.  These models can marginally
resolve the surfaces of the most massive giant molecular clouds ($\sim
70$ pc), and employ a fully radiative sub-resolution model for the
physical and chemical state of GMCs below the resolution limit.  We
consider the main physical processes that determine the thermal
structure of the giant molecular clouds, and thus are able to resolve
their Jeans properties.  As we will show, the high-resolution SPH
galaxy evolution simulations show that the average Jeans mass in a
molecular cloud scales well with the galaxy's star formation rate.

We then aim to understand the cosmic evolution of the Jeans properties
of GMCs, and hence the stellar IMF.  We model the cosmic evolution of
the physical properties of galaxies utilising the analytic methodology
described in \citet{dav12}.  This methodology describes the evolution
of the galaxy star formation rate, stellar mass growth and metal
enrichment while assuming that galaxies live in an equilibrium between
gas inflows, outflows and consumption via star formation.  The results
of the analytic models provide a good match to the evolution of the
physical properties of bona fide hydrodynamic cosmological
simulations, but at a substantially reduced computational cost.  The
main output from these models is the star formation histories for
galaxies of various masses.  Combined with a model for how the average
Jeans properties of molecular clouds vary with the galaxy-wide star
formation rate (as informed from the high-resolution hydrodynamic
galaxy evolution simulations), these models then describe the cosmic
evolution of the IMF under the Jeans Conjecture.

Finally, we combine these results with population synthesis models
\citep[\fsps;][]{con09b,con10a,con10b} in order to explore the
observable properties of stellar populations under a varying IMF.  We
utilise these to compare directly against observations.

\subsection{Hydrodynamic Galaxy Evolution Models}

We simulate the hydrodynamic evolution of a large library of idealised
isolated disc galaxies and galaxy mergers using a modified version of
the publicly available code \gadgettwo\footnote{In practice, we use
  \gadget \ where the principal modifications employed here involve
  processor load balancing.}\citep{spr03a,spr05a,spr05b}.  The
galaxies range in baryonic mass from $\sim 10^{10} \msun$ to $\sim
10^{12} \msunend$, and their physical properties are listed in full in
Table A1 of \citet{nar12a}.  We simulate mergers between these discs
with mass ratio 1:10, 1:3 and 1:1.  Similarly, the initial orbital
configurations are given in \citet{nar12a}.

The disc galaxies are exponential and initialised according to the
\citet{mo98} model.  They are embedded in a live dark matter halo with
a \citet{her90} density profile.  Mergers simply involve two discs
initialised as such.  The halo concentration and virial radius of the
galaxies are motivated by cosmological $N$-body simulations, and are
redshift dependent \citep{bul01,rob06b}.  We simulate galaxies scaled
for \z=0 and \z=3. The gravitational softening lengths are $\sim 100 \ 
\hpc$ for baryons, and $\sim 200 \ \hpc$ for dark matter.

 The \z=0 galaxies are initialised with a 40\% baryonic gas fraction,
 and \z=3 galaxies with an 80\% baryonic gas fraction. These gas
 fractions are potentially larger than those observed
 \citep[e.g.][]{dad10a,tac10,nar12c}. However, because we do not
 include any form of gas replenishment from the intergalactic medium
 (IGM) in these idealised simulations \citep[e.g.][]{mos11a,mos11b},
 we require large initial gas fractions in order to have substantial
 lifetimes of moderately gas-rich galaxies during the evolution of the
 merger.  When galaxies merge, this is necessary to drive a starburst
 \citep{nar10a,hay11,hay12a}. In any case, the exact choice of initial
 gas fraction is not terribly relevant as we principally simulate the
 evolution of idealised galaxies in order to build a sample of
 galaxies with a large dynamic range of physical conditions.  The gas
 is initialised as primordial.  A mass fraction of stars (utilising a
 Salpeter IMF) are assumed to die instantly upon each star formation
 event, and enrich the surrounding ISM with metals assuming
 instantaneous recycling and a yield of 0.02 \citep{spr03a}.

Within the hydrodynamic evolution of the galaxies, the ISM is modeled
as multi-phase, with cold, star-forming clouds embedded in a hotter,
pressure-confining phase \citep{mck77,spr03a}.  These phases can
exchange mass via supernova heating of cold clouds and radiative
cooling of the hotter phase.  Stars form according to a volumetric
\citet{sch59} power-law relation with index 1.5 \citep{ken98a}.  The
normalisation is set to match the local surface density $\Sigma_{\rm
  SFR}-\Sigma_{\rm gas}$ relation \citep{ken98b}.  We note that while
both the slope and normalisation of the Kennicutt-Schmidt star
formation law are the subject of heavy debate \citep[see e.g.][for
  just a few
  examples]{kru07,big08,nar08d,nar08b,kru12a,ost10,ost11,nar11a,nar12b,ken12},
we showed in \citet{nar11b} that the physical conditions in the ISM
are relatively insensitive to these parameters so long as power-law indices
between $\sim 1-2$ are adopted.

\subsection{Molecular Gas ISM Specification}
\label{section:moleculargasspecification}

We determine the physical and chemical properties of the molecular gas
in the SPH galaxy evolution simulations in post-processing.  We
describe the methodology in full in \citet{nar11b}, though summarise
the relevant details here.

\subsubsection{Cloud Chemical and Physical State}

We first project the physical properties of the SPH particles onto an
adaptive mesh using the SPH smoothing kernel.  The base mesh is $5^3$
spanning 200 kpc on a side, and recursively refine in an
oct-subdivision based on the refinement criteria that the relative
density of metals should be less than 0.1, and that the $V$-band
optical depth across a cell be less than unity.  The smallest cells
refined to $\sim 70$ pc across, just resolving massive GMCs.

GMCs within the cells are modeled as spherical and isothermal.  The
\htwo \ fraction is determined by balancing photodissociation rates by
Lyman-Werner band photons against \htwo \ formation rates on dust
grains.  For simplicity, we utilise the analytic prescription of
\citet{kru08,kru09a,kru09b} which assumes equilibrium chemistry.  A
comparison of this model against full radiative transfer models
suggests that this approximation is reasonable for metallicities above
$Z>0.01 Z_{\odot}$, and all our galaxies' metallicities are well above this.

The GMCs are assumed to be of constant density.  In order to treat
GMCs that reside in large cells in the adaptive mesh (and avoid
unphysical conditions), we establish a floor surface density of $100 \ 
\msunpcsq$.  This value is motivated by observations of Local Group
GMCs \citep{bol08,fuk10}.  Generally, in heavily-star forming systems,
the mass-weighted surface density of GMCs is much larger than this,
and the floor value is never reached \citep{nar11b}. For these
``resolved'' GMCs, the cloud is assumed to occupy the entire cubic
volume of the cell that it resides in.  In lower SFR systems, however,
the bulk of the GMCs have $\Sigma_{\rm H2} \approx 100 \ \msunpcsq$.

With the \htwo \ mass and cloud surface density established, and an
assumption regarding the GMC geometry (either spherical[cubic] if the
cloud is unresolved[resolved]), the density is known.  GMCs are known
to be supersonically turbulent \citep[][ and references
  therein.]{mck07}.  We allow for the turbulent compression of gas by
scaling the volumetric densities of the GMCs by a factor
$e^{\sigma_\rho^2/2}$, where numerical simulations show that for
supersonic turbulence:
\begin{equation}
\label{eq:turbulentcompression}
\sigma_\rho^2 \approx {\rm ln}(1+3M_{\rm 1D}^2/4)
\end{equation}
where $M_{\rm 1D}$ is the one-dimensional Mach number \citep[][though
  see \citealt{lem08}]{ost01,pad02}.  As we will show, the temperature
calculation of the GMCs is dependent on the densities.  Because
solving for the density compressions and temperatures simultaneously
for many millions of cells across thousands of galaxy snapshot
realisations is a computationally prohibitive task, we assume the
temperature of the GMC is $10$ K for the sound speed calculation.

\subsubsection{Cloud Thermal State}
\label{section:thermalstate}
The other pertinent variable for calculating the Jeans properties of
the model GMCs is the gas kinetic temperature.  Our model is based on
that laid out by \citet{gol01} and \citet{kru11a}.  The dominant gas
heating terms are by grain photoelectric effect at a rate per H
nucleus $\Gamma_{\rm pe}$, cosmic ray heating at a rate $\Gamma_{\rm
  CR}$, and cooling via either C$_{\rm II}$ fine structure line
cooling, or $^{12}$CO rotational line cooling at a rate $\Lambda_{\rm
  line}$.  Finally, there is an energy exchange between the dust and
gas at a rate $\Psi_{\rm gd}$.  We assume thermal balance between the
gas and dust which gives the following equations that are solved
simultaneously by iterating on the temperatures of the gas and dust:
\begin{eqnarray}
\label{eq:tempequations}
\Gamma_{\rm pe} + \Gamma_{\rm CR} - \Lambda_{\rm line} + \Psi_{\rm gd} = 0\\
\Gamma_{\rm dust} - \Lambda_{\rm dust} - \Psi_{\rm gd} = 0  
\end{eqnarray}

We assume that the photoelectric heating rate is attenuated by half
the mean extinction of the cloud, such that:
\begin{equation}
\Gamma_{\rm pe} = 4 \times 10^{-26}G_0'e^{-N_{\rm H}\sigma_{\rm d}/2} {\rm erg \ s^{-1}}
\end{equation}
where $G_0'$ is the FUV intensity relative to the Solar neighborhood,
and $\sigma_{\rm d}$ is the dust cross section per H atom to UV
photons.  For simplicity, we assume that the $G_0' = 1$ and $\sigma_d
= 1\times10^{-21}$ cm$^{-2}$, though we note that tests that involve
scaling the Habing field by the SFR make little difference on the
thermal properties of our clouds owing to strong shielding
\citep{nar11b}.

The cosmic ray heating rate is given by: 
\begin{equation}
\Gamma_{\rm CR} = \zeta' q_{\rm CR} \ {\rm s^{-1}}
\end{equation}
where $\zeta'$ is the cosmic ray ionisation rate (here, assumed to be
2 $\times 10^{-17} Z' {\rm s}^{-1}$), and $q_{\rm CR}$ is the thermal
energy increase per cosmic ray ionisation.  For \htwo, $q_{\rm CR}
\approx 12.25$ eV \citep[though note that this value is quite
  uncertain; see discussion in Appendix A4 of][]{kru11a}, and for HI,
$q_{\rm CR} \approx 6.5$ eV \citep{dal72}.  

An important aspect of these calculations is that we assume the cosmic
ray heating rate scales linearly with the galaxy-wide SFR, anchored by
an assumed Milky Way SFR of $2 \ \msunyr$ \citep{rob10,cho11}. While the
exact scaling between the cosmic ray ionisation rate and SFR is
unknown, some observational evidence exists suggesting a linear
relationship.  Observations of M82 by the VERITAS group, as well as
$\gamma$-ray observations\footnote{$\gamma$-rays can arise as the result
  of cosmic ray interactions with \htwo, via pion-decay.} of the
Galaxy, Large Magallenic Cloud, NGC 253 and M82 by the FERMI group all
suggest a linear relation between the galaxy-wide SFR and cosmic rays
\citep{acc09,abd10b}.

The dust temperature is calculated via full 3D Monte Carlo radiative
radiative transfer.  We utilise \sunrise, a publicly available dust
radiative transfer simulation package
\citep{jon06a,jon06b,jon10a,jon10b}.  The detailed algorithms are
described in \citet{jon10a}, and we only summarise here.

The sources of light are stellar clusters and an AGN which acts in the
hydrodynamic simulations as a sink particle that accretes according to
a Bondi-Hoyle-Lyttleton parameterisation \citep{bon44}.  The stars
emit a \starburst \ spectrum \citep{lei99,vaz05}, where the ages and
metallicities derive from the SPH calculations.  The AGN emits a
luminosity-dependent SED based on the \citet{hop07} templates that
derive from observations of unreddened type I quasars. In practice,
the AGN does not impact our results significantly \citep[see tests
  by][]{nar11b}.  The radiation traverses the ISM, and is scattered,
absorbed and re-emitted until the dust temperatures converge.  The
dust and radiation field are assumed to be in radiative equilibrium.
The dust mass is set by assuming a constant dust to metals ratio of
$0.4$, comparable to that of the Milky Way \citep{dwe98,vla98,cal08},
and we utilise the \citet{wei01} $R = 3.15$ dust grain models with
\citet{dra07} updates.

For disc galaxies at low-redshift, we assume that young ($<10$ Myr)
stellar clusters are embedded in photodissociation regions and HII
regions for $2-3$ Myr.  In this case, the \starburst \ spectrum is
replaced by SEDs derived from \mappings \ photoionisation calculations
\citep{gro08,jon10a}.  The covering timescale is motivated by
parameter-space surveys by \citet{jon10a} who showed that this
parameter choice produces synthetic SEDs of \z=0 \ disc galaxies
comparable to those observed in the Spitzer Infrared Nearby Galaxy
Survey \citep[SINGS;][]{ken03}.  

For galaxy mergers at \z=0 and all galaxies at \z=3, the stellar
densities become so high that stellar clusters overlap.  In this case,
using \mappings \ no longer makes sense as the super-stellar clusters
(sometime as massive as $\sim 10^8$ \msunend) saturate the \mappings
\ photoionisation models.  In these cases, we assume the cold ISM is a
uniform medium with a volume filling factor of unity.  Observations of
nearby starburst galaxies may support this picture for the cold ISM
structure in heavily star-forming galaxies \citep{dow98,sak99}.
Beyond this, in tests run in the Appendix of \citet{nar11b}, we found
that the thermal and physical properties of the molecular ISM were
generally robust against the choices made for the PDR clearing time or
ISM volume filling factor.

The line cooling is assumed to happen either via CII or CO.  The
fraction of hydrogen where carbon is mostly in the form of CO is well
approximated by \citep{wol10,glo11}:
\begin{equation}
\label{eq:abundance}
f_{\rm CO} = f_{\rm H2} \times e^{-4(0.53-0.045 {\rm
    ln}\frac{G_0'}{n_{\rm H}/{\rm cm^{-3}}}-0.097 {\rm ln}Z')/A_{\rm v}}
\end{equation}
When this fraction is greater than $0.5$, we assume the cooling
happens via CO rotational lines; otherwise, the cooling is dominated
by CII emission.  We assume a constant carbon to \htwo \ abundance of
$1\times10^{-4} \times Z'$ \citep{lee96}, where $Z'$ is the
metallicity with respect to solar.

To determine the cooling rates, we utilise one-dimensional escape
probability calculations \citep{kru07}.  The level populations of the
molecule are assumed to be in statistical equilibrium, and determined
through the rate equations:
\begin{eqnarray}
\label{eq:kmt_stateq}
\sum_l(C_{lu} + \beta_{lu}A_{lu})f_l =
\left[\sum_u(C_{ul}+\beta_{ul}A_{ul})\right]f_u\\
\sum_if_i = 1
\end{eqnarray}
where $C$ are the collisional rates\footnote{The rate coefficients are
  taken from the {\it Leiden Atomic and Molecular Database}
  \citep{sch05}.}, $A_{\rm ul}$ are the Einstein coefficients for
spontaneous emission, $f$ the fractional level populations, and
$\beta_{ul}$ is the escape probability for transition $u\rightarrow
l$.  The rate equations can be rearranged as an eigenvalue problem,
and solved accordingly.

The escape probability, $\beta_{\rm ul}$ can be approximately related
to the line optical depth $\tau_{\rm ul}$ \citep{dra11} via:
\begin{equation}
\label{eq:kmt_beta}
\beta_{ul} \approx \frac{1}{1+0.5\tau_{ul}}
\end{equation} 
and the optical depth is:
\begin{equation}
\label{eq:kmt_tau}
\tau_{ul} =
\frac{g_u}{g_l}\frac{3A_{ul}\lambda_{ul}^3}{16(2\pi)^{3/2}\sigma}QN_{\rm
  H2}f_l\left(1-\frac{f_ug_l}{f_lg_u}\right)
\end{equation}
where $Q$ is the abundance of CO with respect to \htwo, $g_l$ and
$g_u$ are the statistical weights of the levels, $N_{\rm H2}$ is the
column density of \htwo \ through the cloud, $\lambda_{ul}$ is the
wavelength of the transition, and $\sigma$ is the velocity dispersion
in the cloud.  

The 1D velocity dispersion, $\sigma$ is calculated by:
\begin{equation}
\sigma = {\rm max}( \sigma_{\rm cell},\sigma_{\rm vir})
\end{equation}
Here, $\sigma_{\rm cell}$ is the mean square sum of the subgrid
turbulent velocity and the resolved non-thermal velocity dispersion.
The turbulent velocity is calculated from the external pressure from
the hot ISM \citep[][]{rob04}, though we
impose a ceiling of $10$ \kms which is informed by simulations of
turbulent energy driving and dissipation \citep{dib06,jou09,ost11}.
The resolved non-thermal component is calculated as the mean of the
standard deviation of the velocities of the nearest neighbour cells in
the three cardinal directions.  A floor velocity dispersion of
$\sigma_{\rm vir}$ is established (typically for poorly resolved GMCs)
by assuming the GMC is in virial balance with $\alpha_{\rm vir}=1$,
with $\alpha_{\rm vir} \equiv 5 \sigma_{\rm vir}^2 R/(GM)$.

With these elements in place, we then iterate
Equations~\ref{eq:tempequations} and
~\ref{eq:kmt_stateq}-~\ref{eq:kmt_tau} in order to determine the
equilibrium gas temperature.

Finally, we note that our model does not include any effects of X-ray
heating of gas from either massive stars or accreting compact objects
\citep[e.g.][]{hoc10,hoc11}.  This may increase gas heating beyond
the contribution of gas-dust coupling, or cosmic rays in rather
extreme environments.

\subsection{Cosmological Models}
\label{section:cosmology}

As will be discussed in \S~\ref{section:results}, the temperature (and
more generally, the average Jeans mass of a GMC) in a galaxy scale
with the galaxy-wide star formation rate.  In order to understand the
cosmic evolution of the physical properties of galaxies, and therefore
the stellar IMF, we must turn to cosmological galaxy evolution models.

\citet{dav12} presented analytic models that describe the cosmic
evolution of the stellar, gaseous and metal content of galaxies under
the condition that they lie in a slowly evolving (with redshift)
equilibrium state between gas infall from the IGM, gas outflows, and
gas consumption from star formation.  These models compare quite well
to the results from cosmological hydrodynamic simulations
\citep[e.g.][]{opp10}, though have the practical advantage of being
computationally inexpensive.  We therefore utilise these and summarise
the relevant aspects of the model here.

In this model, the star formation rate is determined by the gas infall
rate minus the outflow rate, which is given by the mass loading factor
$\eta$ times the SFR.  The gas infall rate is mitigated by a
preventive feedback parameter $\zeta<1$, while it is augmented by
recycling of winds governed by the recycling parameter
$\alpha_Z$~\citep{fin08}, which is the ratio of the infalling gas
metallicity to the ISM gas-phase metallicity.  The star formation rate
is given by
\begin{equation}\label{eqn:sfrfull}
{\rm SFR} = \frac{\zeta \dot{M}_{\rm grav}}{(1+\eta)(1-\alpha_Z)}.
\end{equation}
Hence there are three parameters in this model, $\eta$, $\zeta$, and
$\alpha_Z$, and their dependences on mass (usually halo) and redshift.
We take $\dot{M}_{\rm grav}$ from \citet{dek09}.

\begin{figure}
\hspace{-1cm}
\includegraphics[scale=0.45,angle=90]{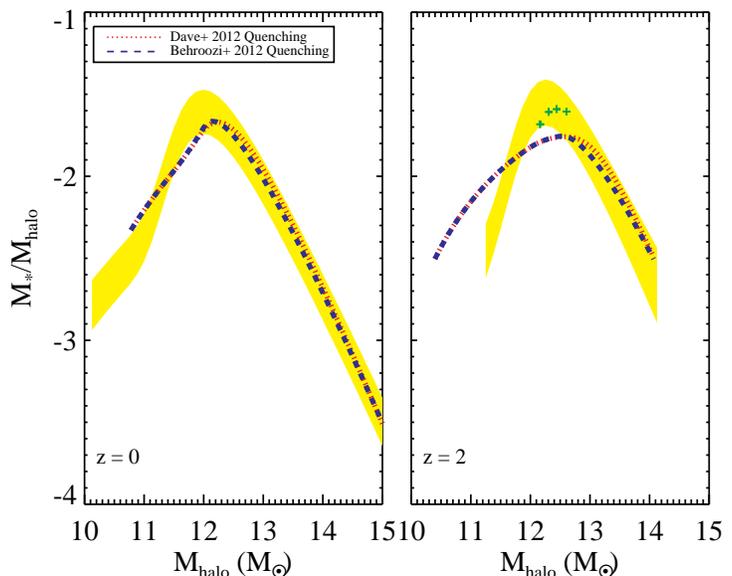}
\caption{M$_*$-M$_{\rm halo}$ relation for model galaxies with
  the two quenching models employed in this work.  The red and blue
  lines show the two different quenching models employed in this work,
  while the yellow band shows observational constraints presented by
  \citet{beh12a,beh12b}.  The left panel shows \zsim 0 galaxies, while
  the right shows galaxies at \zsim 2.  The green crosses in the right
  panel show observational constraints by \citet{wak11}.  At
  low-redshift, galaxies that form in our equilibrium model show
  excellent correspondence with observed galaxies.  Beyond this, the
  different quenching models produce similar galaxies, owing to the
  fact that the bulk of stars are formed prior to quenching.  At
  higher-redshifts, the models may be discrepant from observations at
  low mass, though we note that observations of stellar mass functions
  at \zsim 2 are still relatively uncertain. In any case, the
  correspondence for massive galaxies (the galaxies that concern us in
  this work) is quite reasonable, even at
  high-\z.  \label{figure:mstar_mhalo}}
\end{figure}

Our choices for these three baryon cycling parameters are guided by
numerical simulations and observational results.  We assume
$\eta\propto M_{\rm halo}^{-2/3}$, as advocated by \citet{beh12b} and,
as we show in Figure~\ref{figure:mstar_mhalo}, results in a model
stellar mass to halo mass ratio comparable to what is inferred from
abundance matching.  We take $\alpha_Z$ from a crude parameterization
based on simulations of \citet{dav11b}
\begin{equation}\label{eqn:alphaZ}
(1-\alpha_Z)^{-1}=e^{-Z} \left({{\rm min}(M_{\rm halo}/10^{12} M_\odot,1)}\right)^{2/3}.
\end{equation}
These analytic model choices approximately reproduce simulation
results for key galaxy evolutionary properties such as the star
formation history, metallicities, and gas fractions.

For this work, a key issue is quenching, because as we will show,
low-level star formation in massive galaxies {\it after} quenching is
an important element of our model.  We include a number of sources of
preventive feedback in the model ($\zeta$). 

\begin{equation}
\zeta = \zeta_{\rm photo} \times \zeta_{\rm grav} \times \zeta_{\rm winds} \times \zeta_{\rm quench}
\end{equation}
where $\zeta_{\rm photo}$ represents heating of cold infalling gas by
photoionising radiation, and is principally important at low
masses. Here, we use the parameterisation advocated by \citet{gne00}
and \citet{oka08}, such that $\zeta_{\rm photo} \propto
\left[1+1/3(M/M_\gamma)^{-2}\right]^{-1.5}$ with $M_\gamma$ given in
\citet[][ Figure 1]{dav12}. $\zeta_{\rm grav}$ parameterises the
suppression of gas inflow owing to virial shocks, and is proportional
to $(1+z)^{0.38} \times M_{\rm halo}^{-0.25}$ \citep{fau11}; while 
the simulations that determine this parameterisation for
$\zeta_{\rm grav}$ do not include metal line cooling, results with metal cooling
from \citet{dav11f} confirm this scaling.  $\zeta_{\rm
winds}$ represents the impact of energy input by winds, and scales as
$\sim e^{-\sqrt{M_{\rm halo}}}$ \citep{opp10}.  Generally speaking,
winds and photosuppression of infall do not impact the massive
galaxies that concern us in this study.

$\zeta_{\rm quench}$ represents putative quenching mechanisms
(e.g. feedback from an active galactic nucleus [AGN]) not encompassed
in the other forms of feedback.  For massive galaxy evolution,
$\zeta_{\rm quench}$ tends to have the biggest impact on the SFH which
can impact the exact form of the stellar IMF at \zsim 0 in our model.
Here, we investigate our results in terms of two potential quenching
models.  For the first, we follow \citet{dav12}, and model $\zeta_{\rm
  quench}$ with a similar functional form as $\zeta_{\rm photo}$, but
with a characteristic mass of $M_{\rm q} = 10^{12.3} \msunend$. The
second model appeals to recent work by \citet{beh12b} which utilises
observational constraints on the stellar masses of galaxies to model
the scaling between the cosmic star formation efficiency
(cSFE)\footnote{i.e. The galaxy SFR divided by the halo accretion
  rate. This is to be distinguished from either the mass of stars
  formed divided by the total mass in a molecular cloud
  ($M_*/(M_*+M_{\rm cloud})$), or the SFR/$M_{\rm H2}$ (i.e. an
  inverse time scale), both of which are also commonly referred to as
  the star formation efficiency.} and galaxy halo mass.
\citet{beh12b} find that the cSFE scales as $M_{\rm halo}^{-4/3}$.
When parameterising in terms of the stellar mass-halo mass relation, 
the threshold halo mass beyond which the cSFE becomes inefficient is
$\sim M_{\rm halo} \ga 10^{12} \msun$. Because $\zeta_{\rm grav}$
dominates over $\zeta_{\rm photo}$ and $\zeta_{\rm winds}$ for massive
galaxies, including a quenching term $\zeta_{\rm quench} \propto
M_{\rm halo}^{-1.083}$ for halo masses $M_{\rm halo} > 10^{12} \msun$
results in a cSFE in our models that scales with $M_{\rm
  halo}^{-4/3}$, per the constraints of \citet{beh12b}.

The galaxies that result from the two quenching models are reasonably
similar.  In Figure~\ref{figure:mstar_mhalo}, we plot the \z=0 stellar
mass-halo mass relation for the galaxies in both quenching models.
Owing to the fact that the bulk of the stellar mass is built up prior
to quenching, the two models results in quite similar M$_*$-M$_{\rm
  halo}$ relations.  As we will show, while the low-level star
formation history at late times can affect the present-day stellar IMF
in our model, for plausible choices of the SFH, the results are
qualitatively similar.

This equilibrium model for the cosmological evolution of galaxies
does not account for galaxy mergers.  
While merger-driven starbursts
do little to contribute to the total stellar mass of a
galaxy~\citep[e.g.][]{rod11}, the added stellar content can
affect the observed properties (e.g. the mass to light [$M/L$]
ratio) if the merger is close to equal mass.  Hence we explicitly
include additional growth owing to galaxy mergers.

We estimate the galaxy-galaxy merger rate from the publicly available
model developed in \citet{hop10}.  Briefly, we calculate halo mass
functions and merger rates at a given redshift from the Millenium
simulation \citep{fak10}, and assign galaxies to these haloes
following a standard halo occupation formalism \citep{con09}.  After
correcting for the dynamical friction time-delay between halo-halo
and galaxy-galaxy mergers \citep{boy08}, we then have a galaxy
merger rate as a function of galaxy mass and redshift.  The mapping
between galaxy mass and halo mass is dependent on knowing the galaxy
mass function as a function of redshift.  For this, we utilise
observed galaxy mass functions.  From $z=0-2$, we utilise the mass
functions from \citet{arn07} and \citet{ilb10} as bracketing the
range of observed mass functions. At $z>2$ we utilise mass functions
from \citet{per08} and \citet{mar09}.  The principle effect of
including galaxy mergers is to add to the uncertainty present in
trends as the uncertainty in high-\z \ galaxy mass functions dominates
the uncertainty in these models.  This said, as we discuss later,
the trends presented in this work are robust, and only the normalisation
tends to vary with different mass function choices.  We have run
versions of our model with each of the aforementioned mass functions,
and discuss the range of uncertainty as we present our results.

The combination of simple but realistic star formation histories
for individual galaxies from the equilibrium model plus the merger
history derived from the Millenium simulation allows us to quickly
and effectively model the evolution of the galaxy population.  This
in turn enables us to efficiently examine the impact of IMF variations
parameterized by SFR and metallicity on galaxy properties.

\section{Results}
\label{section:results}

\subsection{Summary of Cloud Thermal State and Scaling of IMF with Galaxy Physical Properties}
\label{section:thermalstatesummary}
Because the methods described in \S~\ref{section:moleculargasspecification}
dictating the thermal state of the clouds are somewhat detailed, we
find it useful to summarise the general trends in cloud temperatures
with the intent of giving the reader an intuition for the different
physical processes that determine the temperature in different
regimes.

Without any sources of heating, the cloud temperatures would cool
to the microwave background temperature.  The dominant sources of
heating are gas-dust coupling when the GMC is at high density ($n\ga
10^{4} \cmthree$), and cosmic ray heating at lower densities ($n\la
10^{2}\cmthree$).

At high densities, the energy exchange between gas and dust becomes
extremely efficient.  In heavily star-forming systems, the gas
temperature is, to first order, set by the dust temperature due to
large fractions of dense gas \citep[e.g.][]{jun09}.  While cosmic rays
can play a significant role in high SFR galaxies
\citep[e.g.][]{pap10a,pap10b}, in these models energy exchange with
dust plays a comparable or greater role in setting the gas temperature
\citep{nar12a}.  Because the dust temperature rises with the SFR
\citep[e.g.][]{nar10b}, the gas temperature does as well. 

In low SFR systems, the mean gas density is relatively low. Here,
cosmic rays determine the gas temperature.  For a galaxy forming stars
at a rate similar to the Galaxy ($\sim 2 \ \msunyrend$), the
resulting temperature is $\sim 10 $ K, comparable to observations of
Galactic clouds \citep{bli07,fuk10}.  At lower SFRs, the cloud
temperature can drop below this value.

The cloud Jeans mass, which scales as $T^{3/2}/n^{1/2}$ therefore
scales, on average, with the galaxy-wide star formation rate.
Physically, this dependence arises from gas-dust coupling at high
densities (combined with increasing dust temperatures with increasing
star formation rates), and the scaling of cosmic ray fluxes with SFRs
at lower densities.  Utilising the information from the $\sim 10^3$
galaxy snapshot realisations modeled here \citep[again, the entire
parameter space surveyed is listed in Table A1 of ][]{nar12a}, we
fit the relationship between the \htwo \ mass-weighted Jeans mass in
our model galaxies and the galaxy SFR.  The \htwo \ mass-weighted
Jeans mass is calculated by weighting the Jeans mass of each cell by
the \htwo \ mass present, and the Jeans mass is calculated utilising
the median gas density in the cell (i.e. the density above which half
the mass resides).  The median density is calculated assuming the gas
has a lognormal density distribution function, with width given by
Equation~\ref{eq:turbulentcompression}.  This results in a relation:
\begin{equation}
\label{eq:mj}
M_{\rm J} \sim {\rm SFR}^{0.3} \msun
\end{equation}
with a normalisation set at a Galactic SFR of 2 \msunyrend.

Under the Jeans conjecture, the scaling of the average Jeans mass of
molecular clouds has implications for the stellar IMF.  We assume the
IMF has a similar basic shape in all galaxies, and takes the form of a
Kroupa broken power-law with a a low-mass slope of 0.3, and high-mass
slope of -1.3 (where the IMF slope specifies the slope of the
log$_{\rm 10}$(dN/dlogM)-log$_{\rm 10}$(M) relation).  For Milky Way
conditions, the turnover mass, $M_{\rm c}$, is set to $0.5 $ \msunend,
and varies with the SFR for other galaxies.  Thus, for
IMF variations, we have:
\begin{equation}
\label{eq:imf}
M_{\rm c} = 0.5 \times \left[\frac{{\rm SFR}}{{\rm SFR_{\rm MW}}}\right]^{0.3}
\end{equation}

\begin{figure}
\hspace{-1cm}
\includegraphics[angle=90,scale=0.4]{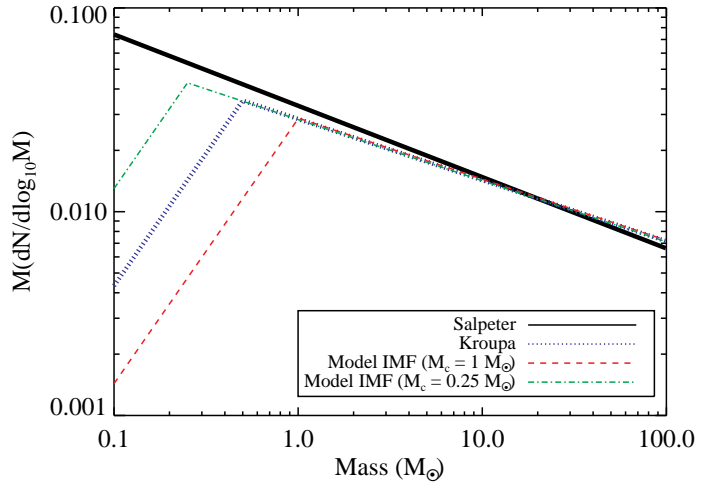}
\caption{Example stellar initial mass functions.  The black solid line
  denotes a standard \citet{sal55} IMF, whereas the other three show
  \citet{kro02} broken power-law forms.  The blue dashed line is the
  typical Galactic Kroupa IMF with turnover mass $M_{\rm c} = 0.5 \ 
  \msunend$.  The green dash-dot line represents a bottom-heavy IMF (with
  $M_{\rm c} = 0.25 \ \msun$), while the red dashed line is a top-heavy
  IMF (with $M_{\rm c} = 1 \ \msun$).  At solar metallicities, the
  bottom-heavy IMF may represent a galaxy forming stars at $\sim 0.2 \ 
  \msunyrend$, while the top-heavy IMF is characteristic of a galaxy
  forming stars at $\sim20-25 \  \msunyrend.$ An important point is that
  the shape of the IMF does not change under the model assumption of
  the Jeans conjecture: only the turnover mass.  The IMFs are
  normalised arbitrarily by the integral $\int{\rm M(dN/dlog_{\rm
      10}M) \ dM}$.\label{figure:imf_example}}
\end{figure}

Under the Jeans conjecture, the IMF therefore varies with the
conditions in the ISM.  Physically, the IMF tends toward top-heavy
in heavily star-forming galaxies due to efficient energy exchange
between gas and dust in dense environments.  Conversely, the IMF
tends toward bottom-heavy in poorly star-forming galaxies owing to
decreased cosmic ray fluxes.  It is important to note that the {\it
shape} of the IMF does not change in this model - only the turnover
mass. To explicitly show this point, in Figure~\ref{figure:imf_example},
along with a standard Salpeter and Kroupa IMF, we show example IMFs
for a low-SFR bottom-heavy system ($M_{\rm c} = 0.25 \ \msunend$), and
a high-SFR top-heavy system ($M_{\rm c} = 1 \ \msunend$).

As we will see, Equation~\ref{eq:imf} combined with the results from
\S~\ref{section:cosmology} prepare us to analyse the cosmic evolution
of the IMF under the assumption of the Jeans conjecture.  Finally, we
note that formally, the rise in Jeans masses in high-SFR galaxies can
be modulated in very low metallicity environments owing to a
metallicity-dependent dust-gas energy exchange rate \citep{kru11a}.
Typically, heavily star-forming systems do not remain metal-poor in
our model for very long, however, and any metallicity-dependence in
Equation~\ref{eq:imf} has minimal impact on the IMF in massive
galaxies at \zsim 0 under the Jeans conjecture.  Future work will
discuss the star formation properties of low-metallicity, heavily
star-forming systems at high-redshift.

\subsection{The Meandering of the IMF over Cosmic Time}
\label{subsection:cosmicevolution}

\begin{figure*}
\includegraphics[scale=0.8]{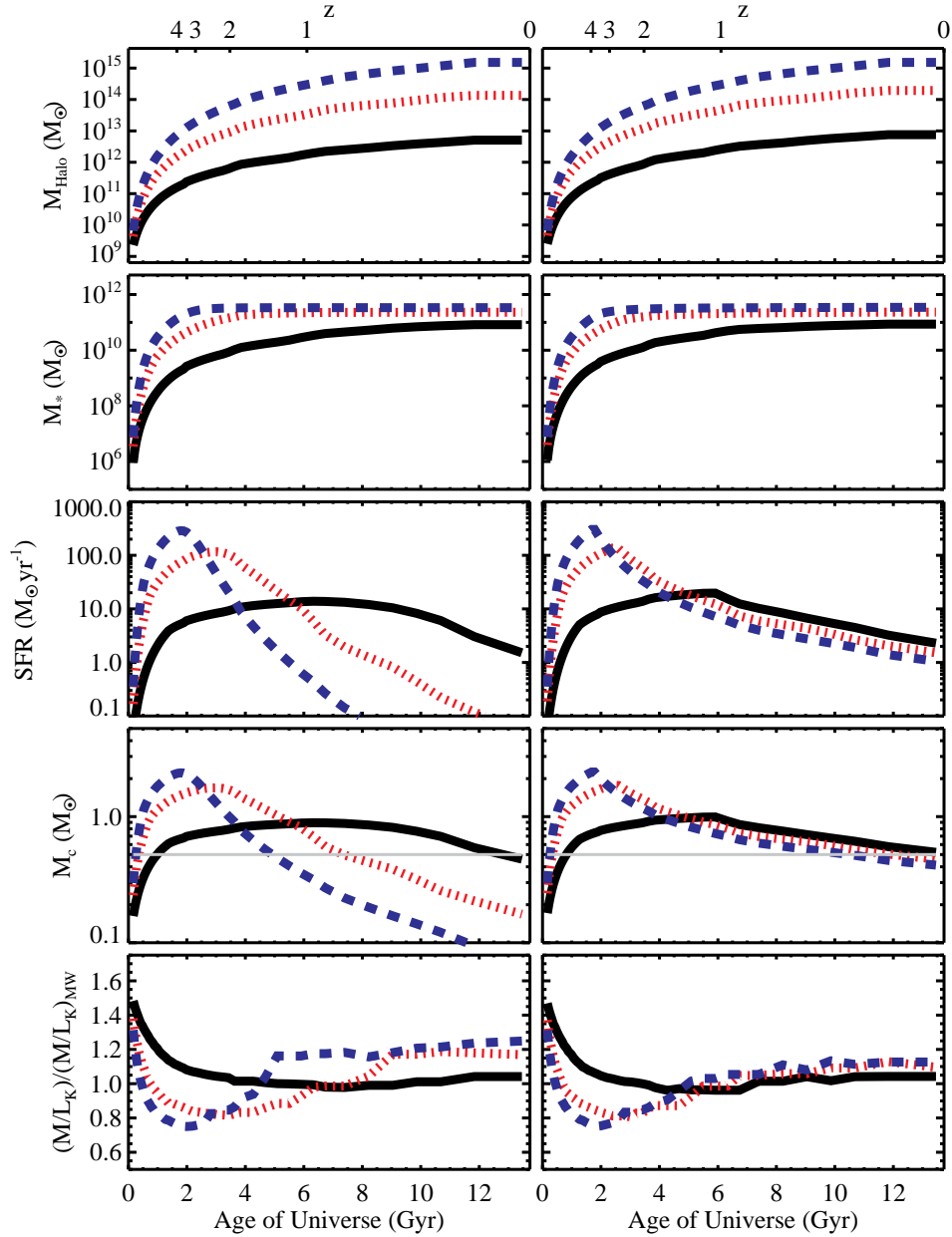}
\caption{Cosmic evolution of model galaxy halo masses, $M_*$, SFRs,
  IMF turnover masses, and $K$-band mass to light ratios for three
  galaxy masses. The left column shows the quenching model from
  \citet{dav12}, while the right column shows the quenching model
  designed to match the constraints of \citet{beh12b}.  The galaxies
  are picked to have the same final stellar mass for each quenching
  model. The thin grey line in the fourth panel denotes $M_{\rm c}=0.5$
  \msunend, the turnover mass for a Galactic Kroupa IMF.  The mass to
  light ratios are normalised by the expected values for a Kroupa IMF
  to assist in comparing with observations \citep{con12b}. The small
  features in the $M/L$ ratio owe to the discretization of the SFH
  into individual bursts, as well as galaxy mergers (see text for
  details). The IMF for a massive galaxy may vary over cosmic time
  from top-heavy to bottom-heavy as follows.  During epochs of heavy
  star formation, the Jeans masses of clouds increase owing to gas in
  clouds heated by warm dust in dense environments.  These galaxies
  may exhibit top-heavy IMFs under the Jeans conjecture.  Galaxies of
  increasing mass peak in their cosmic star formation history at
  earlier times, and undergo long periods of star formation at rates
  below the Milky Way's.  Decreased cosmic ray fluxes during low SFR
  periods result in cloud Jeans masses lower than typical Galactic
  values, and hence an excess of low-mass stars in their IMFs at late
  times.  \label{figure:m2l_time}}
\end{figure*}

In Figure~\ref{figure:m2l_time}, we plot the cosmic evolution of the
halo mass, $M_*$, SFR, IMF characteristic mass (c.f.
Figure~\ref{figure:imf_example}), and K-band mass to light ratio
(normalised by what one would see for a standard Kroupa IMF) for three
model galaxies for both quenching models.  The model galaxies are
chosen to have the same final stellar mass. We defer discussion of the
mass to light ratio evolution until \S~\ref{section:bottomheavy}.

The characteristic star formation history of galaxies is, to first
order, determined by the gas accretion rate from the IGM which is
dictated by gravity.  Fits to models from various groups all find
accretion rates dependent on both the halo mass, as well as redshift
via $\dot{M}_{\rm grav} \propto M_{\rm halo}^{0.05-0.15} \times
z^{2-2.5}$ \citep[e.g.][]{dek09,fak10,fau11}.

In this picture, more massive galaxies on average have higher peak
SFRs, and peak at earlier times~\citep[e.g.][]{noe07b,noe07a}.  As an example,
while a (\z=0) Milky Way sized-halo peaks in SFR between $\z=0-1$,
and has a peak SFR of $\sim 10-20$ \msunyrend, a galaxy in a ($\z=0$)
$~\sim 10^{14} \  \msun$ halo peaks at $\sim 200-300$ \msunyr at
$\zsim 2-3$.  Modulo metallicity effects (which only play an important
role for the most part at redshifts $\z\ga 4$), the evolution in the
turnover mass in the IMF broadly follows the star formation history.
Turning to the third panel in Figure~\ref{figure:m2l_time}, we now
highlight the grey horizontal line which denotes the typical Kroupa
turnover mass of $M_{\rm c} = 0.5 \  \msun$ in the Milky Way.

Massive galaxies that undergo heavy SFR periods at early times see an
increase in their average cloud temperatures\footnote{The cloud
  densities of course increase as well, but the Jeans mass goes as
  $M_{\rm J} \sim T^{3/2}/n^{1/2}$.}, and hence an increase in the
characteristic turnover mass in their stellar IMFs at early times
(e.g. top-heavy/bottom-light IMFs).  In \citet{nar12b}, we showed that
the top-heavy IMF in high-\z \ star-forming galaxies that results from
a Jeans model can alleviate tensions associated with inferred SFRs
from high-\z \ galaxies.

As the SFRs of massive galaxies decline due to a combination of gas
exhaustion and cosmic expansion of the Universe, so do their cosmic
ray fluxes.  Below \zsim 2, when the cosmic microwave background (CMB)
temperature decreases below $T_{\rm CMB} \la 10$ K, the temperatures
of the GMCs may decrease below the typical $\sim 10$ K value seen in
Milky Way clouds.  Consequently, the cloud Jeans masses drop and the
IMF becomes bottom-heavy.  For the most massive galaxies we consider,
the transition between top-heavy and bottom-heavy (defined as when the
turnover mass in the IMF transitions from above $0.5 \ \msun$ to
below) happens at $\zsim 1-2$.

Under the Jeans conjecture, then, the IMF meanders over cosmic time,
generally guided by the galaxy-wide SFR.  The exact form of the
meandering is galaxy mass-dependent. For example, the most massive
galaxies ($M_{\rm halo} \sim 10^{14} \ \msun$ at $\zsim 0$) peak in
their SFR, and hence their IMF characteristic mass, at \zsim 2-4.
They then go through a bottom-heavy phase for roughly half a Hubble
time. It is important to note, though, that the bulk of the stars that
form during the galaxy's lifetime do so during the top-heavy phase.
It is only {\it newly} formed stellar clusters that will exhibit a
bottom-heavy IMF in massive galaxies at \zsim 0 under the assumption
of the Jeans conjecture.

 Galaxies that reside in Milky Way-sized haloes at \zsim 0 peak in
 their SFRs so late that they never undergo a bottom-heavy phase, and
 are only weakly top-heavy at late times ($\z \la 1$).  Interestingly,
 galaxies that are much less massive than the Milky Way never have
 large ($\ga 2$ \msunyrend) star formation rates, and thus have
 bottom-heavy IMFs for the bulk of their lives in this model.

It is important to note short-term fluctuations in the cosmic
evolution of the star formation history from a given galaxy can
cause deviations in the IMF from the trends presented in
Figure~\ref{figure:m2l_time} and discussed in this section.  For
example, a gas-rich major merger that drives an intense burst of
star formation can temporarily drive up gas temperatures, and hence
the cloud Jeans masses.  This said, under the Jeans conjecture, on
average the IMF trends will follow those shown in
Figure~\ref{figure:m2l_time}.  We now expand upon the trends discussed
here and compare more specifically to inferred measurements of the
IMF from observations.

\subsection{The Mass-to-Light Ratio in \z=0 Massive Ellipticals}
\label{section:bottomheavy}

As discussed in the Introduction, various circumstantial arguments
favour a bottom-light IMF in rapidly star-forming galaxies at high
redshifts.  Such arguments are detailed more thoroughly in
\citet{nar12b}, where we consider the implications of the Jeans
mass conjecture on high-$z$ galaxies.  In \S\ref{sec:starbursts}
we discuss arguments for and against the interpretation of a
bottom-light (or top-heavy) IMF in massive high-$z$ galaxies.  Here,
we consider the implications of our varying IMF for the descendents
of massive high-$z$ galaxies, which today are expected to be
early-type ``red and dead" ellipticals.

Observationally, a number of works have found that the K-band
mass-to-light ratio ($M/L_K$) increases with stellar mass more quickly
than expected for a standard Kroupa or Chabrier IMF
\citep[e.g. ][]{tre10,cap12b,con12a,con12b}. Observationally,
this is parameterised as an increase in $M/L_K$ with galaxy velocity
dispersion, $\sigma$.

 What happens in our Jeans conjecture model?  Referring again to
 Figure~\ref{figure:m2l_time}, a trend in the characteristic IMF mass
 at \zsim 0 is apparent with increasing galaxy mass. While a Milky Way
 mass galaxy peaks in its SFR around \zsim 1-2, and only settles to an
 IMF with $M_{\rm c} \approx 0.5 \ \msun$ around \zsim 0, more massive
 galaxies peak in their SFR at earlier times.  These galaxies exhaust
 their fuel rapidly, are unable to replenish their gas owing to strong
 preventive feedback, and settle into a low SFR mode (perhaps in an
 episodic fashion) for up to $\sim$ half a Hubble time.  By present
 epoch, the only signature of the massive stars that formed during the
 peak star formation event are their stellar remnants.

We employ stellar population synthesis models to derive the stellar
mass to light ratio from our model galaxies.  We utilise \fsps, a
flexible, publicly available stellar population synthesis code
described in \citet{con09b,con10a} and \citet{con10b}.  For a given
galaxy, we utilise the star formation history and cosmic evolution
of the IMF to evaluate the properties of the \z=0 stellar populations.
We model a given galaxy by discretizing the star formation history
into a series of bursts.  We run \fsps \ assuming the star formation
history is a single burst at each redshift, with the corresponding
$M_{\rm c}$ derived from our Jeans model.  We then sum the mass to
light ratios, weighting by the stellar masses, of the galaxy over
its cosmic history to determine the mass to light ratio at a given
redshift.

We now refer to the bottom panel of Figure~\ref{figure:m2l_time}.
Here, we show $M/L_K$ normalised by what is expected from a standard
Kroupa IMF.  The masses include the contribution from stellar remnants
(e.g. white dwarfs and neutron stars). As is evident, a galaxy
residing in a Milky Way sized halo at $\zsim 0$ has a $M/L_K$ ratio
comparable to what is expected from a standard Kroupa IMF.  Galaxies
of increasing mass have increasingly large $M/L_K$ ratios.

\begin{figure}
\hspace{-1.2cm}
\includegraphics[angle=90,scale=0.425]{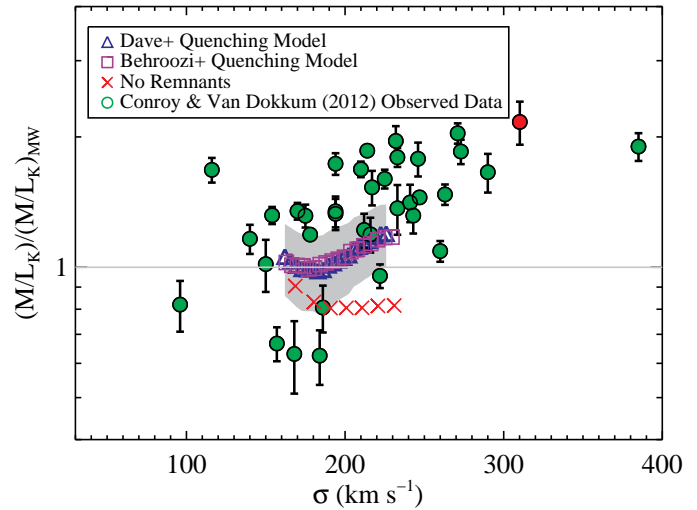}
\caption{Stellar mass to ($K$-band) light ratios as a function of
  galaxy stellar velocity dispersion for simulated galaxies (blue
  triangles and purple squares), and observed galaxies \citep[green
    and red circles with error bars;][]{con12b}.  The blue triangles
  have SFHs with the \citet{dav12} quenching model imposed, while the
  purple squares have a quenching model designed to match constraints
  on the cosmic star formation efficiency-M$_{\rm halo}$ relation from
  \citet{beh12b}. The $M/L_K$ ratios are normalised by what is
  expected from a Kroupa IMF.  The grey shaded region denotes a 0.2
  dex uncertainty in the normalisation of the models (see text for
  details).  Galaxies of increasing mass have larger SFRs at earlier
  epochs.  Due to stellar remnants that remain from massive stellar
  evolution, the observed $M/L$ ratios are higher at present epoch.
  See text for details.  \label{figure:m2l_charlie}}
\end{figure}

This trend is more clearly seen in Figure~\ref{figure:m2l_charlie},
where we examine the $\z=0$ mass to light ratios in our model
galaxies against stellar velocity dispersion (serving as a proxy
for galaxy mass).  To remain consistent with the way observations
are presented, we normalise $M/L_K$ by a Milky Way ratio (here, a
Kroupa IMF with $M_{\rm c} = 0.5 \ \msunend$; C. Conroy, private
communication).  The velocity dispersion is calculated utilising
fits to the dynamical mass versus velocity dispersion from Sloan
Digitized Sky Survey (SDSS) galaxies presented in \citet{van11b}.

The blue triangles and purple squares in
Figure~\ref{figure:m2l_charlie} denote the individual galaxy
realisations from the cosmological simulations.  The overall
normalisation of the simulated points is not very robust against our
model assumptions.  Variables such as the assumed Galactic SFR (which
dictates when decreased cosmic ray fluxes will allow cloud
temperatures to drop below $\sim 10$ K) and the assumed stellar mass
function at high-redshift (which can affect the galaxy merger rate;
c.f.  \S~\ref{section:cosmology}) can slightly change the
normalisation of the $M/L_K$ ratio as a function of mass.  The typical
amount varying these parameters within reasonable bounds changes the
$M/L_K$ ratios is $\sim$0.2 dex and this region of uncertainty is
highlighted by the grey shaded region (only noted for the Dav\'e
quenching model).  The magnitude of the trend in $M/L_K$ with $\sigma$
depends on the exact form the SFH (and hence the quenching model).
This said, for plausible star formation histories, a general trend of
increasing $M/L_K$ ratio with $\sigma$ is evident.  

The points with error bars show data from the observational work of
\citet{con12b}.  The green points are a sample of nearby early type
galaxies (including the bulge of M31), and the red point is stacked
data from 4 galaxies in the Virgo cluster.  The data show increasing
mass to light ratios with increasing stellar velocity dispersion.
Similar trends have been noted by \citet{tre10,smi12} and
\citet{cap12b} as well. The model results are qualitatively in
agreement with observational data.  They indicate that galaxies with
$\sigma$ typical of the Milky Way represent a minimum in the
mass-to-light ratio in our Jeans conjecture IMF relative to a Kroupa
IMF.  Galaxies of increasing mass (or velocity dispersion) show
increasing mass-to-light ratios.

The $M/L_K$ ratio increases with galaxy velocity dispersion due to the
increased contribution from stellar remnants in our models.  Galaxies
of increasing mass have undergone increasingly bottom-light phases at
earlier times, and thus have larger contributions to their stellar
masses at present epoch by stellar remnants.  



Galaxies at lower velocity dispersions, in this model, also tend to
have increased mass-to-light ratios.  In these galaxies, the stellar
population is actually bottom-heavy.  This result comes from the
protracted epoch of low star formation associated with massive
galaxies that populate the low-mass end of the IMF at late times,
while in low-mass galaxies (which are typically still star-forming
today) the low SFR's and metallicities result in a bottom-heavy IMF at
all times.

When comparing to observations, two salient points arise. First, while
the $M/L$ ratio increases with increasing galaxy mass in our model,
this is due to stellar remnants from the top-heavy phase, not an
increasingly bottom-heavy IMF.  This is seen more explicitly via the
red crosses in Figure~\ref{figure:m2l_charlie}, where we plot the
$M/L$ ratio of our models with the effects of stellar remnants turned
off in the SPS calculations.  The shape of $M/L$ with $\sigma$ for the
case of no remnants owes to a competition between lack of stellar
remnants in high-mass galaxies with increasingly bottom-heavy IMF's at
late times.  

 Still, the correspondence of our fiducial model with observations is
 reasonable.  Many of the observational studies aimed at understanding
 the IMF in galaxies infer the IMF indirectly by constraining the
 $M/L$ ratio.  If the Jeans hypothesis is correct, then the observed
 incrase in $M/L_K$ ratio with galaxy mass owes to a prior top-heavy
 phase, rather than bottom-heavy phase.  The two scenarios are
 degenerate in terms of observed $M/L$ ratios.  This said,
 measurements of gravity-sensitive stellar absorption features by
 groups such as \citet{con12b} and others directly probe the stellar
 IMF within the central regions of galaxies.  If indeed the IMF is
 bottom-heavy as these studies imply, this may provide difficulty for
 the Jeans model.

Second, in detail, this model cannot attain the $\sim\times 2$
increase in $M/L_K$ relative to Kroupa as seen in some galaxies,
including the stack of Virgo ellipticals (red point).  It may be
possible to tweak model parameters to achieve this.  For instance, we
have assumed that the IMF form is always Kroupa-like; if instead the
IMF also had a shift to a shallower slope, this would result in more
massive stars that would disappear by today, and would increase the
relative contribution of the low-mass stars created during the low-SFR
phases of evolution, when the IMF was more bottom-heavy.

Galaxy mergers moderately enhance the $M/L$ ratio in massive galaxies.
This is because in our model, low mass galaxies have a bona fide
(slightly) bottom-heavy IMF as they never achieve significant periods
of star formation above SFR$_{\rm MW}$.  The massive galaxies of
course have a large $M/L$ due to the presence of remnants.  As a
result, mergers with low mass galaxies can add more low mass stars and
slightly increase the observed $M/L_{K}$ ratio.  This said, we do not
account for the increased SFR driven by galaxy mergers.
Qualitatively, one would expect to observe a top-heavy IMF in an
ongoing merger due to increased star formation rates and hence Jeans
masses (via increased gas temperatures).  A complete accounting of all
these effects would require implementing a full evolutionary IMF model
within a cosmologically-based galaxy formation model that includes the
effects of mergers self-consistently; we leave this for future work.

\section{Discussion} \label{section:discussion}

In \citet{nar12b} we discussed the implications of an IMF following
the Jeans conjecture on the properties of high-redshift star forming
galaxies.  We showed that this IMF goes significantly towards
alleviating a number of difficulties in understanding the evolution
of the SFR-$M_*$ relation and sub-millimetre galaxies, and has
interesting implications for the star formation law.

In this paper we consider the implications of the Jeans conjecture
on the evolution of passive, massive galaxies.  While our main focus
is their mass-to-light ratios, there are a number of other implications
for massive galaxies that we now briefly consider.

\subsection{Galaxy Star Formation and Chemical Enrichment History}

For a typical massive galaxy in this model, the star formation history
peaks at high-redshift, and once the halo mass exceeds the quenching
mass $M_{\rm quench}$ its star formation begins to become strongly
suppressed~\citep[see Figure~1 of][]{dav12}.  The physical mechanism
for this is purported to be AGN feedback~\citep[e.g.][]{cro06},
although our model is purely parameteric.  The resulting star
formation histories for different quenching models are shown in
Figure~\ref{figure:m2l_time}: more massive galaxies have earlier and
higher peaks of star formation, consistent with stellar population
data for early-type galaxies~\citep{tho05b}, and reminiscent of the
observationally-derived ``staged" galaxy formation
model~\citep{noe07b,noe07a}.  Importantly, the star formation rate
remains finite, albeit small, even after quenching has kicked in.
This SFH integral is important for driving top-heavy IMFs in galaxies
during high star formation rate epochs, and the observed $M/L$ ratio
in present-epoch galaxies.

The star formation histories presented here are broadly consistent
with observations.  Recent studies find that low levels of star
formation and dense molecular gas is widespread among $z\sim 0$ early
type galaxies \citep{you11,xu10,cro11,dav11e,cro12}, Additionally,
observations by the ATLAS$^{\rm 3D}$ collaboration appear to
tentatively suggest that ellipticals have star formation histories
consistent with those presented here, with more massive galaxies
typically having older stellar populations \citep[][]{mcd12}.  In this
sense, that our model suggests that more massive galaxies peak in
their SFH at increasingly early times, followed by low-level residual
star formation may be reasonable.

While early-type galaxies clearly undergo some small levels of star
formation, there is significant variance in the amount of star
formation among galaxies, suggesting a duty cycle for this ``frosting"
of star formation.  In our simple model, this would result in a
less bottom-heavy $M/L_K$, because the actual star formation rates
would be higher (albeit over shorter periods of time).  However,
this would not result in a change in the {\it trend} of $M/L_K$ 
unless there was a systematic variation in the duty cycle with
$\sigma$.

Another issue relates to the chemical enrichment history of
ellipticals, which can be significantly impacted by changes in the
IMF.  Galaxies with short star formation time scales are expected to
have enhanced [$\alpha/$Fe] ratios. This is because when star
formation is burst-like, the enrichment is dominated by type II
supernovae, thus the stars will be enhanced in elements that are
products of the $\alpha$ process in stellar evolution.  On the other
hand, when star formation occurs over an extended period, there is
time for type Ia supernovae to contribute Fe to the chemical makeup of
the galaxy.  Observationally, more massive galaxies appear to be more
$\alpha$-enhanced \citep{tra00}.  \citet{arr10} showed that more
top-heavy mass functions will increase the $[\alpha]$/Fe ratios in
galaxies, and that top-heavy models may be needed to match the
observed enrichment trends.  Without introducing a full chemical
evolution model (and its associated uncertainties) to these
simulations, however, being more quantitative is currently infeasible.

In summary, observations show both evidence for ongoing star formation
in early-type galaxies \citep{cro11}, older stellar populations on
average for more massive galaxies \citep{tho05b,mcd12,pac12}, and
increasing [$\alpha$/Fe] ratios in early types \citep{tra00},
suggesting that the typical star formation histories calculated by
our cosmological galaxy evolution model may not be unreasonable.
However, current uncertainties in the detailed SFH and chemical
enrichment histories of early-type galaxies are such that these
issues cannot definitely confirm or rule out a Jeans conjecture
origin for the bottom-heavy IMF in ellipticals.

\subsection{Is the IMF in Rapidly Star-forming Galaxies Top-Heavy?}

\label{sec:starbursts}

The Jeans conjecture IMF model predicts that ellipticals undergo an
early bottom-light/top-heavy phase during its peak star formation rate
epoch.  \citet{nar12b} have suggested that this may go some distance
toward alleviating a number of difficulties in understanding the
evolution of high-$z$ galaxies.  In the Jeans model, newly formed
clusters in a present epoch massive galaxy will form in a bottom-heavy
fashion, though the bulk of the stars formed over cosmic time will
have done so in a bottom-light fashion.

Other works have investigated the potential for a top-heavy IMF in
massive starbursts.  For example, \citet{wei11b} have investigated the
effects of crowding of cloud cores in GMCs, and found that the
integrated galactic IMF may become top-heavy when the galaxy SFR is $>
10$ \msunyr \citep[see also ][]{kro11}.

There is, alternatively, a general class of models that aims to
understand the origin of the bottom-heavy IMF in ellipticals by making
the IMF more bottom-heavy during phases of heavy starbursts.
\citet{bat09} and \citet{kru11d} examined the role of radiative
pressure in setting the characteristic mass of the IMF.  Krumholz
showed that if the characteristic mass is set by radiative feedback
from young protostars, then the IMF will scale very weakly with
pressure such that in high-pressure systems (i.e. the heavily
star-forming galaxies that serve as the progenitors for local
early-types) a bottom-heavy IMF results.  Similarly, \citet{hop12b}
showed that if the IMF is set by the sonic mass in gas, rather than
the Jeans mass, then starbursts will show an excess of low-mass stars.
While neither model directly examines the observable properties of a
$\zsim 0$ early-type galaxy population, it is likely that they too
would show an excess of low-mass stars in accord with observations.
In this sense, whether the IMF is top-heavy or bottom-heavy in
starburst systems may be a discriminant in general classes of IMF
models.  It is therefore worth examining claims of top-heavy IMFs in
high SFR systems critically.

A number of studies have noted a mismatch between the observed cosmic
evolution of stellar mass density and cosmic star formation rate
density such that the inferred SFR density produces a factor $\sim 2$
more stars at $\zsim 0$ than are measured \citep{hop06d,wil08c}.
While \citet{nar12b} showed that an IMF that varies with the Jeans
masses in galaxies could go some distance toward reconciling these
differences (producing only a mismatch in the stellar masses at $\zsim
0$ of a factor 1.3 instead of 2), some observational studies
either fail to find such a mismatch \citep[e.g. ][]{sob12}, or propose
non-IMF based solutions to the apparent discrepancy
\citep{red09,sta12}.

Another commonly invoked candidate for a top-heavy IMF in starburst
galaxies is the most heavily star-forming galaxy population in the
Universe, the \zsim 2 Submillimetre-selected population.  The
arguments are primarily theoretical.  Simply, most standard galaxy
formation models have been unable to account for the full observed
distribution of SMGs without resorting to a top-heavy IMF.  The most
extreme example was reported by \citet{bau05} who suggested a flat IMF
may be necessary to reconcile the observed SMG number counts,
Lyman-break galaxy population and present-day stellar mass function.
However, recent numerical simulations by \citet{hay10} and
\citet{hay12b} show that if one accounts for a combination of effects,
including small scale dust obscuration in highly-resolved simulations,
the contribution of isolated galaxies, and the contribution of galaxy
pairs, the observed SMG number counts may be accounted for even under
the assumption of a standard Kroupa IMF.

\citet{van08} utilised the fact that the luminosity and colours of a
galaxy may evolve differently, depending on the IMF, and found that a
bottom-light/top-heavy IMF best fit the evolution of the $M/L_{\rm B}$
and $U-V$ colours for $0<z<1$ early-type galaxies in clusters.
However, when accounting for potential variations in population
synthesis models, frosting of young stellar populations in the
galaxies \citep{tra08}, and the structural evolution of galaxies
\citep{hol10}, \citet{van12} found that the same observations could be
consistent with a Salpeter IMF.

Other evidence for bottom-light IMFs in heavily star-forming galaxies
come based on an analysis of H$\alpha$ equivalent widths and optical
colours of $\sim 33,000$ galaxies in the GAMA survey \citep{gun11}.
Similarly, comparisons of H$\alpha$ to UV luminosity ratios have
suggested more top-heavy IMFs in more heavily star-forming galaxies
\citep{lee09,meu09}, though at least some of these results may be
explained via stochasticity in the formation of massive stars
\citep{fum11} or variable star formation histories \citep{wei12}.

In general, the purported IMF variations toward top-heavy are all
fairly mild, and while it is tempting to solve a number of apparent
problems at high-\z \ with a mild shift in the IMF towards
bottom-light/top-heavy, there could be other explanations for each of
these observations.  This said, it is interesting that all inferred
IMF variations in heavily star-forming systems go in the direction of
being more top-heavy/bottom-light. Whether heavily star-forming
galaxies have a top-heavy or bottom-heavy IMF should be a direct
discriminant between general classes of IMF theory.

In principle, one could use remnants to potentially distinguish
whether the IMF for massive galaxies has gone through a top-heavy
phase at early times.  If galaxies undergo a top-heavy phase, then
their dynamical masses will have a larger contribution from stellar
remnants.  This may be consistent with observed fractions of low mass
X-ray binaries in ellipticals \citep[e.g.][]{kim09a}.  On the other
hand, if the bulk of the stellar mass is built up in a bottom-heavy
phase, then the contribution from remnants to the dynamical mass will
be minimal.  Offsets between direct $M/L$ measurements from stellar
spectral features (which measure the masses of living stars) and
dynamical modeling that includes the effect of remnants may
potentially provide a constraint on this model.

The main idea of this paper is that the mass-to-light ratios in
today's ellipticals is consistent with a top-heavy IMF at earlier
times in heavily star-forming systems under a Jeans model for the IMF.
This is due to increased stellar remnants.  Newly formed clusters in
fact form in a bottom-heavy fashion in this model, though do not
comprise the majority of the stellar mass.  If the IMF is truly
bottom-heavy in massive galaxies (as recent results from stellar
absorption line spectra suggest), then this poses a challenge for the
Jeans conjecture, and a bottom-heavy IMF at all epochs would more
straightforwardly solve the problem (of course, at the cost of
potentially exacerbating possible issues with high-redshift galaxies).
Observations that distinguish whether the IMF varies in heavily
star-forming systems, or as a function of formation age \citep[e.g
][]{zar12} may prove to be a valuable model discriminant.

\citet{wei13a} present a toy model in which they allow the IMF switch
from bottom-light to bottom-heavy early during the SFH of a
representative galaxy.  This sort of model is able to both reproduce
the observed chemical enrichment of massive galaxies, as well as the
potential bottom-heavy nature of their IMF.  While this model does not
self-consistently explain {\it why} the IMF should switch rapidly from
bottom-light to bottom-heavy, it does give some insight as to what the
timing for IMF changes should look like during a galaxy's SFH.  If the
IMF of heavily star-forming galaxies is truly bottom-light, and of
their remnants, bottom-heavy, then the \citet{wei13a} toy model serves
as a guide for constraining when and how rapidly the IMF needs to
change to match observed constraints.

\subsection{The Validity of the Jeans Conjecture}

The most basic arguments surrounding the Jeans conjecture is that it
is not entirely clear what scale one should average over in order to
derive the physical conditions in the cloud.  In this work, we assume
that the IMF scales with global cloud properties, on scales of $\sim
70$ pc. If one were to choose significantly smaller scales, as an
example, it is possible that the temperature and density would not vary
in a manner that kept the Jeans mass constant. A number of works have
posited that a reasonable choice for scale may be one where the
isothermality of the cloud is broken.  \citet{lar05} suggested that
this is when the gas becomes energetically coupled to the dust, though
other works have argued that the isothermality may be broken due to
radiative feedback \citep{bat09,kru11d}.  

While we cannot resolve these scales, what likely matters most is not
the physical conditions in GMCs as a whole, but rather the clumps that
form stellar clusters.  Our model follows the typical enhancements and
decrements of the Jeans mass on cloud-wide scales given the physical
conditions in a galaxy.  So long as the physical conditions in
star-forming clumps within a GMC scale in a self-similar manner with
the global physical properties of the molecular cloud, then the
galactic environment may indeed play a role in setting the
fragmentation scale that impacts the distribution of core masses. 

From a simulation standpoint, some tentative evidence exists that
suggests a close connection between the cloud Jeans masses and IMF.
Work by \citet{bat05} and \citet{kle07} showed that the mass spectrum
of collapsed objects varies with the Jeans properties of the simulated
cloud, and that conditions comparable to starbursts may have turnover
masses as large as $\mc \approx 15 \ \msunend$.  This said, the
degree of fragmentation is dependent on the exact form of the
equation of state, and the scaling of the turnover mass may depend
on the numerical resolution for isothermal simulations.

Perhaps the most convincing observational evidence for the Jeans
conjecture is that it is reasonably well-established that the stellar
IMF corresponds well with the cloud core mass function in some GMCs
\citep{and96,joh01,mot01,nut07,and11}.  If these cores are the result
of clump fragmentation that lead to star formation
\citep{eva99,lad03,ken12}, then a relationship between the
characteristic fragmentation scale and the IMF is natural. There is
additionally some evidence from the Taurus star-forming region and
Orion Nebula Cluster that suggest variations in the IMF with cloud
density in a manner that is consistent with a Jeans scaling
\citep{bri02,luh03,luh04,dar12}, though the dynamic range of physical
conditions on global cloud scales in the Galaxy is limited enough that
unambiguous evidence for an IMF that varies with cloud Jeans masses is
not yet present.  ALMA observations will allow for better measurements
of the relationship between GMC properties and core mass functions
over a range of physical conditions which may shed light on the
potential relationship between the stellar IMF and cloud Jeans mass.

Whether the IMF varies with the Jeans properties of clouds, or at all
with gas physical conditions remains an open question.  While some
observations show mild variations, unambiguous evidence for
systematic variations are lacking \citep{bas10}.  The Jeans
conjecture equally enjoys its share of tentative observational
evidence, as well as contends against cogent theoretical
arguments. In this work, we remain agnostic as to the validity of the
Jeans conjecture, but rather simply aim to understand the potential
consequences of an IMF that scales with global cloud properties.

\section{Summary}
\label{section:summary}

Utilising a combination of hydrodynamic galaxy evolution simulations,
a radiative model for the interstellar medium, and cosmological models
for the star formation histories of galaxies, we investigate the
cosmic evolution of the stellar initial mass function under the
assumption that the IMF varies with the Jeans masses of molecular
clouds in galaxies.

The IMF can be both top heavy and bottom heavy (at different times)
for massive galaxies.  At early times, the ISM is warm owing to the
heating of dust by stars, and energy exchange between gas and dust.
This drives Jeans masses in molecular clouds up, and, under the Jeans
conjecture, results in a top-heavy IMF.  

As galaxies evolve toward present epoch, their SFRs drop below Milky
Way values, and a lack of cosmic rays allows the temperature of the
ISM to cool below the $\sim 10$ K characteristic of Galactic clouds.
At these late times, the typical Jeans masses of clouds decreases, and
the IMF tends toward an excess of low-mass stars (bottom-heavy).
Newly formed stellar clusters in prsent-epoch massive galaxies will
form with a bottom-heavy IMF under the Jeans Conjecture.  This said,
the majority of stars formed will have done so in a top-heavy phase,
and are thus challenged by stellar absorption-line measurements that
suggest that massive galaxies at \zsim 0 have IMFs that are
increasingly bottom-heavy with $\sigma$.

More massive galaxies form a larger number of stellar remnants during
their top-heavy phases, and thus have increased $M/L$ ratios.  After
convolving our models with stellar population synthesis calcuations,
we find that the resultant $M/L$ ratio for massive ellipticals
increases with increasing galaxy mass. Given plausible
galaxy star formation histories, a Jeans model for IMF variations
provides a reasonable match to observed $M/L - \sigma$ relations for
observed elliptical galaxies at \zsim 0.

In short, by exploring the ansatz that the stellar IMF varies with the
Jeans properties of clouds galaxies, we conclude that a Jeans model
for IMF variations in galaxies is able to simultaneously reproduce the
observed $M/L$ ratios of present epoch massive galaxies, while
motivating top-heavy IMFs at early times that may help alleviate some
tensions in high-\z \ galaxy evolution.  The Jeans model suggests that
increased $M/L$ ratios in present-epoch massive galaxies owes to
stellar remnants.  If the IMF in massive galaxies is truly
bottom-heavy, as gravity-sensitive spectral absorption features may
indicate, then the Jeans conjecture may not be valid.

 We remain agnostic as to the underlying Jeans assumption, as well as
 regarding inferred IMF variations at both low and high-\z.  Improving
 constraints on the IMF in starbursting systems and confirming
 preliminary trends of varying IMF with star cluster age would seem to
 provide viable paths towards discriminating the Jeans conjecture
 model versus models that invoke bottom-heavy IMFs throughout the
 lifetime of today's massive galaxies.

\section*{Acknowledgements}
DN thanks Peter Behroozi, Charlie Conroy, Phil Hopkins, Mark Krumholz,
Richard McDermid, Daniel Stark, Rachel Somerville, Jesse van de Sande,
and Pieter van Dokkum for interesting and helpful conversations.  We
additionally thank Peter Behroozi, Charlie Conroy and David Wake for
providing the data from their papers in digital format.  We thank the
referee for numerous suggestions that improved this work.  This work
benefited from work done and conversations had at the Aspen Center for
Physics.  DN acknowledges support from the NSF via grant AST-1009452.
RD is supported by the National Science Foundation under grant numbers
AST-0847667 and AST-0907998, and by NASA under grant number
NNX12AH86G.  Computing resources were obtained through grant number
DMS-0619881 from the National Science Foundation.

\newpage

\end{document}